\documentclass[sigconf,anonymous=false]{acmart}
\AtBeginDocument{%
  }

\setcopyright{acmlicensed}
\copyrightyear{2025} 
\acmYear{2025} 
\setcopyright{cc}
\setcctype{by}
\acmConference[RecSys '25]{Proceedings of the Nineteenth ACM Conference on Recommender Systems}{September 22--26, 2025}{Prague, Czech Republic}
\acmBooktitle{Proceedings of the Nineteenth ACM Conference on Recommender Systems (RecSys '25), September 22--26, 2025, Prague, Czech Republic}\acmDOI{10.1145/3705328.3748016}

\acmISBN{}

\usepackage[ruled,vlined,linesnumbered]{algorithm2e}
\usepackage{color,soul}
\usepackage{colortbl}

\newcommand\rdw{$RP^{3}_{\smash{\beta}}$}
\newcommand\drdw{D-RDW}

\begin{document}

\title[Diversity-Driven Random Walks]{D-RDW: Diversity-Driven Random Walks\\for News Recommender Systems}


\author{Runze Li}
\email{runze.li@uzh.ch}
\orcid{0009-0001-7284-753X}
\affiliation{%
  \institution{Department of Informatics,\\University of Zurich}
  \streetaddress{Binzmühlestrasse 14}
  \city{Zurich}
  \country{Switzerland}
  \postcode{CH-8050}
}

\author{Lucien Heitz}
\email{heitz@ifi.uzh.ch}
\orcid{0000-0001-7987-8446}
\affiliation{%
  \institution{Department of Informatics,\\University of Zurich}
  \streetaddress{Binzmühlestrasse 14}
  \city{Zurich}
  \country{Switzerland}
  \postcode{CH-8050}
}

\author{Oana Inel}
\email{inel@ifi.uzh.ch}
\orcid{0000-0003-4691-6586}
\affiliation{%
  \institution{Department of Informatics,\\University of Zurich}
  \streetaddress{Binzmühlestrasse 14}
  \city{Zurich}
  \country{Switzerland}
  \postcode{CH-8050}
}

\author{Abraham Bernstein}
\email{bernstein@ifi.uzh.ch}
\orcid{0000-0002-0128-4602}
\affiliation{%
  \institution{Department of Informatics,\\University of Zurich}
  \streetaddress{Binzmühlestrasse 14}
  \city{Zurich}
  \country{Switzerland}
  \postcode{CH-8050}
}

\renewcommand{\shortauthors}{Li, Heitz, Inel, Bernstein}

\begin{abstract}
    This paper introduces \textbf{D}iversity-Driven \textbf{R}an\textbf{d}om \textbf{W}alks (\drdw), a lightweight algorithm and re-ranking technique that generates diverse news recommendations. 
    \drdw{} is a societal recommender, 
    which combines the diversification capabilities of the traditional random walk algorithms with customizable target distributions of news article properties.
    In doing so, our model provides a transparent approach for editors to incorporate norms and values into the recommendation process.
    \drdw{} shows enhanced performance across key diversity metrics that consider the articles' sentiment and political party mentions when compared to state-of-the-art neural models.
    Furthermore, \drdw{} proves to be more computationally efficient than existing approaches. 
\end{abstract}

\ccsdesc[500]{Information systems~Information retrieval diversity}
\ccsdesc[500]{Information systems~Recommender systems}
\ccsdesc[500]{Theory of computation~Random walks and Markov chains}

\maketitle

\section{Introduction}
\label{sec:introduction}
In news recommender systems (NRSs), diversity is one of the most important criteria for assessing the societal impact of these systems~\cite{bernstein_et_al:DagMan.9.1.43,vrijenhoek2022radio,sargeant2022spotlight,tintarev2024measuring}.
This is particularly true for recommending political articles, as a diverse selection of news can impact opinion formation and social deliberation~\cite{helberger2019democratic,heitz2022benefits,heitz2023deliberative}.
When designing NRSs, however, diversity optimization is predominantly achieved in a post-processing step~\cite{vargas2014coverage,petersen2021post}.
As such, it plays only a secondary role; it is limited to re-ranking items from the candidate list of the model, instead of being able to access the entire item pool~\cite{wan2023processing,heitz2023classification}. 
To address the limitations of re-ranking approaches, this paper introduces \emph{\textbf{D}iversity-Driven \textbf{R}an\textbf{d}om \textbf{W}alks} (\drdw) that provides diversity optimization at the model stage.\newline

\drdw{} is an extension of the random walk algorithms \rdw~\cite{christoffel2015blockbusters,paudel2021random}.
In the past, \rdw{} was shown to outperform neural models across a wide range of accuracy and diversity metrics in the movie and e-commerce domain~\cite{christoffel2015blockbusters,ferrari2019we}.
While state-of-the-art neural models can account for diversity through re-ranking, the benefits of using random walks for news recommendations are that they allow for
1) an explainable recommendation approach (i.e., they are not a black box model),
2) an easily extendable approach to include heuristics of editors and journalists, and 
3) a lightweight and cost-efficient approach that is highly scalable and parallelizable.
\drdw{} achieves this by introducing a new step in the random walk pipeline that enforces a so-called normative target distribution (NTD) of article properties in the recommendation list.

As a secondary contribution, we demonstrate the effectiveness of enforcing NTD to achieve higher diversity by creating separate re-ranking strategies for existing neural models.
When speaking of normativity in the context of NRSs, we follow the definition of~\citet{vrijenhoek2023normalize}, which refers to the process of operationalizing and including societal values in the optimization and evaluation of NRSs (e.g., optimizing news feeds for source or political viewpoint diversity and providing equal exposure for minority and majority parties).
In doing so, we offer a first diversity-optimizing model and re-ranker for a \textit{deliberative recommender} according to the normative NRSs framework of~\citet{helberger2019democratic}.

\section{Related Work}
\label{sec:background}
Diversity has increasingly been recognized as a crucial aspect of assessing the quality of news recommendations~\cite{bauer2024values}, by looking at aspects such as topics, viewpoints, and political mentions, among others~\cite{bernstein_et_al:DagMan.9.1.43,rao2013taxonomy,bachmann2022defining,baden2017conceptualizing}.
NRSs also take into account other forms of diversity, including those related to sources, people, events, semantics, sentiments, authors, and temporal aspects~\cite{chakraborty2017optimizing,wu2020sentirec}. 
In recent years, however, the research community has begun to recognize the critical role that NRSs play in democratic societies and has proposed four different models from the normative framework of democracy to consider exposure diversity as a societal goal~\cite{helberger2018exposure,helberger2019democratic}.
These normative frameworks, however, have not yet been implemented into any recommendation model.

\begin{figure*}
    \includegraphics[width=175mm]{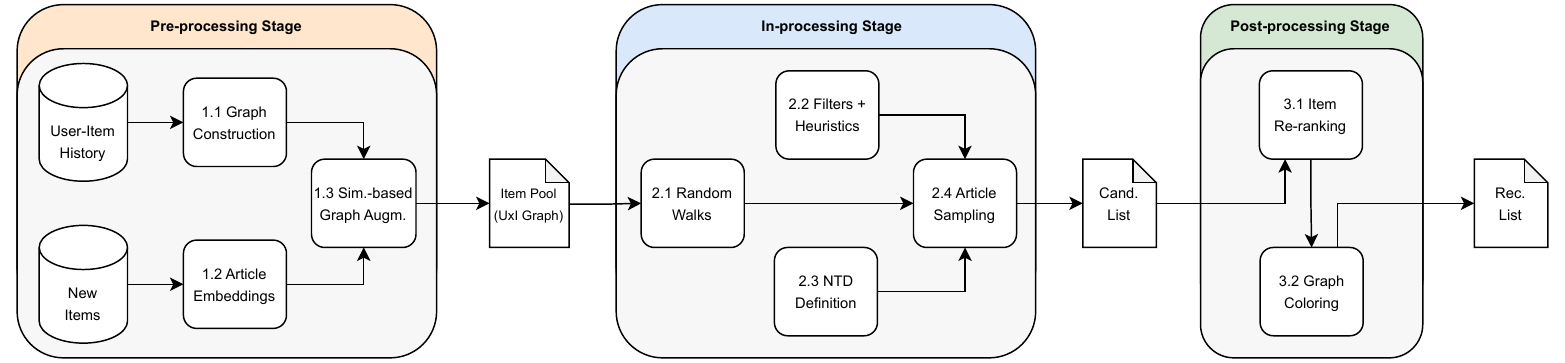}
    \caption{The \drdw{} pipeline consists of three stages: 1) a pre-processing stage for graph construction, 2) an in-processing stage for generating a candidate list, and 3) a post-processing step for creating the recommendation list.}
    \label{fig:random_walk_pipeline}
    \Description{Description}
\end{figure*}

Looking at models, several approaches in the literature focused on generating such diversified recommendations by using random walk-based models, i.e., models that simulate user navigation in an item-user interaction graph as a probabilistic transition~\cite{baluja2008video,eksombatchai2018pixie,de2015investigation,wang2018rndm,yu2019recommendation,liu2008eigenrank}.
\citet{liu2012solving} introduced a collaborative filtering method based on directed random walks that amplifies the influence of users with small degrees to reach a more diverse set of recommendations in the movie domain.

Christoffel et al. ~\cite{christoffel2015blockbusters,paudel2016updatable} created an improved variant, \rdw{}, that showed the capability of random walks to optimize for both accuracy as well as diversity.
~\citet{paudel2021random} then proposed Random Walk with Erasure (RWE-D), where they successfully leveraged random walks to diversify recommendations within a left-right political spectrum.
These approaches, however, do not offer any direct control to editors or journalists when it comes to specifying the precise dimension of diversity (be that, e.g., source or viewpoint diversity).
The underlying algorithm only looks at user interactions and has no information on the article content.
\drdw{} addresses this issue with NTD-based filtering by accessing annotated news articles, which we will outline in the next section.

\section{Diversity-Driven Random Walks}
\label{sec:drdw}
\drdw{} follows the recommendation pipeline shown in Figure \ref{fig:random_walk_pipeline}.
%
It consists of three main steps:
1) a pre-processing stage for graph construction based on user-item interactions,
2) an in-processing stage for running random walks and applying filter criteria to enforce a target normative item distribution, and
3) a post-processing stage for re-ranking the candidate list to achieve a homogeneous distribution of articles across the news feed.\footnote{A reference implementation of \drdw{} is available online in our repository: \url{https://github.com/Informfully/Experiments}}

\paragraph{\textbf{Stage 1: Graph Construction and Augmentation}}
The random walk algorithm uses a bipartite graph with user and item nodes~\cite{christoffel2015blockbusters,paudel2021random}, with edges representing user-item interactions (i.e., clicks).
To address the cold start problem, unconnected nodes (i.e., articles not yet read by any user) are integrated by selecting the most similar items already in the graph\footnote{Items are matched based on the semantic similarity of the article text. Similarity is calculated with an off-the-shelf model for embeddings: \url{https://huggingface.co/sentence-transformers/all-mpnet-base-v2}} and creating edges between the new item and \emph{all users} connected to the $3$ most similar articles that are part of the graph.

\paragraph{\textbf{Stage 2: Random Walks and Normative Distributions}}
Recommendations are generated by following four steps:
1) running random walks on the graph with $3$ hops, 
2) applying heuristic rule for removing items included in the history of the target user,
3) defining a normative target distribution (NTD), 
and
4) sampling articles to find a set of items that satisfies the NTD (see Equation~\ref{eq:sample}) using a customizable objective function. 

The novel element of \drdw{} is the article sampling in Algorithm~\ref{alg:drdw} on Line~\ref{line:distribution} to solve the constraint satisfaction problem of the NTD. 
For example, let $I$ be an all-ones vector in a matrix form $I^\intercal X = 20$, with $X$ being a vector where each element $x_i \in \{0,1\}$ represents whether item $i$ is selected and $20$ is the target size of the recommendation.
This equation ensures that exactly 20 items are selected.
We then define a binary indicator matrix for each category.
Take $P$, where $P_i = 1$ represents the political item $i$ and 
let $S_i = 1$ be an item $i$ with a positive sentiment. 
In this example, our NTD includes $15$ political articles and $10$ articles with a positive sentiment.
By combining these linear equations into their matrix form, we arrive at Equation~\ref{eq:sample}:
\begin{equation}
    \begin{pmatrix}
        I^\intercal \\
        P^\intercal \\
        S^\intercal \\
     \end{pmatrix} X = 
     \begin{pmatrix}
        20 \\
        15 \\
        10 \\
     \end{pmatrix}
     \label{eq:sample}
\end{equation}
The vector \( X \) is restricted to binary values ($0$ or $1$), making it a binary integer programming problem. 
As multiple solutions for \( X \) may satisfy these linear equations, users can specify an optional sampling objective function.
By default, we use the random walk transition probability as the objective, but this can be replaced by any \textit{numeric} item feature (e.g., recency).
Let $C$ represent a vector of item attributes that the sampling algorithm aims to maximize in the final item selection.
When $C$ is defined as the item recency vector, the sampling goal is to select items that satisfy the linear constraints while maximizing overall recency, formalized as $max \sum_i c_i \cdot x_i$.

In cases where the sampling process is unable to achieve an ideal \textit{full set} of $20$ items that satisfies all constraints, we iteratively reduce the target size by one to identify the maximum-sized \textit{subset} that meets both the ratios of the NTDs and the adjusted target size.

\paragraph{\textbf{Stage 3: Item Re-ranking and Graph Coloring}}
%
\drdw{} can sort the candidate list by random walk transition probability or prediction score.
If the candidate list is not a \textit{full set}, the remaining slots are filled with randomly selected articles (Line \ref{line:sampling}). 
%
An optional step of applying a graph coloring algorithm~\cite{kierstead2000simple} to the candidate list is included.
This ensures a homogeneous category distribution across the recommendation list.
We note that the graph coloring algorithm is not interfering with the diversity of the recommended items and the offline experiments presented here.
Instead, it is intended for the visualization of the recommended items in online studies (cf.~\cite{heitz2024informfully}).
We address this matter as future work.

\paragraph{\textbf{Normativity and Explainability}}
With \drdw, our goal is to provide an algorithm for NRSs that is both norm-aware and easy to explain.
In terms of operationalizing norms, we base our algorithm on the work of \citet{helberger2019democratic}, where four types of democratic recommenders are proposed.
They are each defined by the different distributions of, e.g., topics, viewpoints, and emotions in the recommendations.
In our approach, 
we use article category as a proxy for modeling the distribution of topics, 
party mentions as a proxy for viewpoints, and sentiment as a proxy for emotion.
Based on this, \drdw{} implements what is referred to as the \textit{deliberative} approach, focusing on an equal representation of viewpoints (party mentions) and diversity of emotions (article sentiment).\footnote{Please see Section~\ref{sec:experiment} for the precise numbers used in the underlying NTD.}

By using a customizable NTD for item filtering, \drdw{} provides a way to tweak the output of random walks that takes into account article properties.
As such, it goes beyond existing algorithms whose parameters only allowed for changes to the output based on the graph-related properties (e.g., number of hops~\cite{christoffel2015blockbusters} or node degree~\cite{paudel2016updatable}).
NTDs with a focus on category, party mentions, and sentiment present properties that are easy to quantify and explain.
They are non-technical terms that editors in the newsrooms are familiar with.
In addition, the explainability capabilities of our algorithm would provide support for conducting user studies in the future to assess the effectiveness of normative-driven nudges on the news consumption of users~\cite{modre2023value,heitz2024idea}. 

\begin{algorithm}

    \DontPrintSemicolon
    \caption{Diversity-Driven Random Walks}
    \label{alg:drdw}
    
    \SetKwInOut{KwIn}{Input} \label{alg:in}
    \SetKwInOut{KwOut}{Output} \label{alg:out}
    \KwIn{$graph, articleAttributes, targetDimensions$\\ $targetDistribution, targetSize, filterCriteria$\\ $samplingObjective, maxHops$}
    \KwOut{$recommendations$}
    
    $recommendations = []$ \\
    $currentHop = 3$ \\
    
    \While{$True$}{
        $candidates = newHop(graph, currentHop)$\\ \label{line:hop}
        $candidates = filterHeuristics(candidates,$\\ 
                        \qquad $articleAttributes,filteringCriteria)$\\
        $recommendations = sampling(candidates,$\\ \label{line:distribution}
                        \qquad $articleAttributes,targetDimensions,$\\ 
                        \qquad $targetDistribution,targetSize,$\\ 
                        \qquad $samplingObjective)$

        \If{$length(recommendations) >= targetSize \: \vee $\\ 
            \qquad $ currentHop >= maxHops$}{
               $\textbf{break}$
        }
        
        $currentHop = currentHop + 2$
    }

    \If{$length(recommendations) < targetSize$}{ \label{line:sampling}
           $recommendations = addRandomArticles(candidates,$\\ 
                        \qquad $articleAttributes,targetDimensions,$\\ 
                        \qquad $targetDistribution,targetSize,$\\ 
                        \qquad $samplingObjective)$
    }

    $recommendations = rankArticles(recommendations,$\\
                        \qquad $targetSize,rankingObjectives$)
    
    \Return $recommendations$
    
\end{algorithm}

\section{Experiment}
\label{sec:experiment}
We evaluate \drdw{} on the RecSys 2024 Challenge benchmarking Ekstra Bladet News Recommendation Dataset (EB-NeRD), specifically EB-NeRD small~\cite{kruse2024eb}.
Recommendations are calculated using the test set with $15,339$ users. 
In our analysis, we compare the performance of D-RDW with the neural models LSTUR~\cite{an2019lstur}, NPA~\cite{wu2019npa}, and NRMS~\cite{wu2019nrms}. 
To optimize for diversity, we re-ranked the output of these models using three customized approaches: greedy KL (G-KL)~\cite{steck2018calibrated}, PM-2~\cite{dang2012diversity}, and MMR~\cite{carbonell1998use}.\footnote{G-KL and PM-2 use the same NTD as \drdw.}
Additionally, \rdw{}, random walks with erasure (RWE-D), and a randomized list (Random) are added as baseline algorithms.\footnote{We used $3$ hops and for augmenting the graphs of the random walk models, we connected each cold item with the users of their $3$ most similar items in the graph.} 
%
We experimented with different numbers of epochs and ratios for the sampling strategy of negative and positive items for our baseline neural models.
However, similar to previous experiments with EB-NeRD that used a subset of our baselines~\cite{kruse2024eb}, we were unable to significantly improve AUC beyond $0.6$ and used the original hyperparameters instead.\footnote{For details, please see the official implementation found on the Microsoft Recommenders GitHub repository: \url{https://github.com/recommenders-team/recommenders}}

For our NTD, we consider sentiment and political party distributions in this experiment.
Sentiment ranges from negative ($-1.0$) to positive ($1.0$).\footnote{We used the XLM-roBERTa model: \url{https://huggingface.co/cardiffnlp/twitter-xlm-roberta-base-sentiment}}
The articles are put into $4$ discretized buckets with value ranges of $[-1,-0.5)$ (20\%), $[-0.5,0)$ (30\%), $[0,0.5)$ (30\%), $[0.5,1]$ (20\%).\footnote{Percentages indicate the share of each bucket in the final recommendation list (e.g., 20\% of all articles in the user's feed should have a positive sentiment of $0.5$ or higher).}
The political party is based on mentions.\footnote{EB-NeRD covers predominantly news from Denmark. We therefore model NTD according to the Danish political landscape, by considering the governing and supporting parties and the opposition parties. Wikipedia overview: \url{https://en.wikipedia.org/wiki/Folketing}}
NTD has buckets for articles that mention: 
1) government parties (incl. supporting parties, 15\%), 
2) opposition parties (15\%), 
3) government \textit{and} opposition parties (15\%),
4) independent/foreign parties (\textit{must} mention at least one independent/foreign party, \textit{may} mention a government or oppositions party, 15\%), and 
5) no political parties (40\%).
We purposefully measure category performance while not including this dimension in our proposed NTD.
The reason for doing so is to assess the impact of the added NTD sampling step for random walks when comparing category performance to, e.g., party performance that is part of our NTD.
Finally, we would like to emphasize again that this NTD configuration is only one of many possible ways of operationalizing deliberative diversity.

%
%

We assess the model performance using the area under the ROC curve (AUC)~\cite{provost2001robust} for test impressions.
We follow the example of~\citet{vrijenhoek2022radio} and calculate the top $20$ predictions, as diversity metrics require a large recommendation list to calculate divergence metrics.
However, the EB-NeRD small test set has a median of only $11$ impression items per user session (with a median of $1$ clicked item per session).
We, therefore, rank and subsequently select items from among the entire test pool of $4,400+$ articles to generate the recommendation lists.
This new setting makes assessing accuracy metrics challenging, as the expected value of the default scenario (e.g., predicting top $1$ out of $11$ impressions for EB-NeRD, cf.~\cite{kruse2024recsys,kruse2024eb}) is vastly different from our setting (top $1$ out of $4,400$ items).
We, therefore, only report on AUC for the base models. 

\begin{table*}

    \caption{Overview of the diversity scores for the top 20 news recommendations and AUC for item prediction scores of the entire article pool. Results closest to the NTV are highlighted in \colorbox{green!17}{green}, second closest in \colorbox{blue!17}{blue}, and third closest in \colorbox{orange!17}{orange}.}
    \scalebox{0.735}{
       
        \begin{tabular}{lc|cccccc|cccccc|cc|c}

            \toprule
            
            \textbf{Model} &
                \parbox{1.65cm}{\centering{\textbf{Re-ranking}}} &
                \parbox{0.975cm}{\centering{\textbf{Activ.}}} &
                \parbox{0.975cm}{\centering{\textbf{Cat. Calib.}}} &
                \parbox{0.975cm}{\centering{\textbf{Comp. Calib.}}} &
                \parbox{0.975cm}{\centering{\textbf{Frag.}}} &
                \parbox{0.975cm}{\centering{\textbf{Alt. Voices}}} &
                \parbox{0.975cm}{\centering{\textbf{Repr.}}} &
                \parbox{0.975cm}{\centering{\textbf{Cat. Gini}}} &
                \parbox{0.975cm}{\centering{\textbf{Sent. Gini}}} &
                \parbox{0.975cm}{\centering{\textbf{Party Gini}}} &
                \parbox{0.975cm}{\centering{C\textbf{at. ILD}}} &
                \parbox{0.975cm}{\centering{\textbf{Sent. ILD}}} &
                \parbox{0.975cm}{\centering{\textbf{Party ILD}}} &
                \parbox{0.975cm}{\centering{\textbf{Train. Cost}}} &
                \parbox{0.975cm}{\centering{\textbf{Rec. Cost}}} &
                \parbox{0.975cm}{\centering{\textbf{AUC}}} \\
            
            \midrule
            
            NTV &
                 &
                1.000 &
                0.000 &
                0.000 &
                0.000 &
                0.000 &
                1.000 &
                0.000 &
                0.133 &
                0.250 &
                1.000 &
                0.779 &
                0.789 &
                0.000 &
                0.000 &
                1.000 \\

            \midrule
        
            LSTUR~\cite{an2019lstur} &
                 &
                0.190 &
                \cellcolor{orange!17}0.433 &
                0.257 &
                0.610 &
                0.063 &
                0.372 &
                0.832 &
                0.645 &
                0.874 &
                0.740 &
                0.552 &
                0.343 &
                351.12 &
                69.84 &
                \cellcolor{blue!17}0.564 \\

            NPA~\cite{wu2019npa} &
                 &
                0.202 &
                0.468 &
                0.266 &
                0.619 &
                \cellcolor{orange!17}0.061 &
                0.365 &
                0.826 & 
                0.605 &
                0.877 &
                0.753 & 
                0.580 &
                0.337 &
                523.05 &
                51.25 &
                \cellcolor{orange!17}0.554 \\

            NRMS~\cite{wu2019nrms} &
                 &
                0.204 &
                0.509 &
                0.285 &
                0.632 &
                \cellcolor{blue!17}0.060 &
                0.362 &
                \cellcolor{blue!17}0.791 &
                0.544 &
                0.880 &
                \cellcolor{orange!17}0.794 &
                0.624 &
                0.326 &
                326.55 &
                63.23 & 
                0.549 \\

            \midrule

            LSTUR &
                G-KL &
                0.282 &
                0.444 &
                0.240 &
                0.612 &
                0.104 &
                0.546 &
                0.828 &
                \cellcolor{green!17}0.133 &
                \cellcolor{green!17}0.250 &
                0.754 &
                \cellcolor{green!17}0.779 &
                \cellcolor{green!17}0.789 &
                351.12 &
                30.15 &
                 \\

            &
                PM-2 &
                0.284 &
                0.440 &
                0.244 &
                0.604 &
                0.115 &
                0.546 &
                0.819 &
                \cellcolor{blue!17}0.150 &
                \cellcolor{green!17}0.250 &
                0.766 &
                \cellcolor{blue!17}0.776 &
                \cellcolor{green!17}0.789 &
                351.12 &
                23.86 &
                 \\
                
            &
                MMR &
                0.295 &
                \cellcolor{orange!17}0.433 &
                \cellcolor{orange!17}0.235 &
                0.613 &
                0.118 &
                \cellcolor{green!17}0.575 &
                0.823 &
                0.226 &
                \cellcolor{blue!17}0.270 &
                0.759 &
                0.762 &
                \cellcolor{blue!17}0.807 &
                351.12 &
                52.36 &
                 \\

            \midrule

            NPA &
                G-KL &
                0.292 &
                0.469 &
                0.237 &
                0.589 &
                0.108 &
                0.545 &
                0.810 &
                \cellcolor{green!17}0.133 &
                \cellcolor{green!17}0.250 &
                0.775 &
                \cellcolor{green!17}0.779 &
                \cellcolor{green!17}0.789 &
                523.05 &
                30.15 &
                 \\

             &
                PM-2 &
                0.302 &
                0.471 &
                0.239 &
                \cellcolor{orange!17}0.576 &
                0.122 &
                0.547 &
                0.806 &
                \cellcolor{blue!17}0.150 &
                \cellcolor{green!17}0.250 &
                0.783 &
                \cellcolor{blue!17}0.776 &
                \cellcolor{green!17}0.789 &
                523.05 &
                23.86 &
                 \\
            
            &
                MMR &
                0.314 &
                0.463 &
                0.236 &
                0.598 &
                0.110 &
                \cellcolor{green!17}0.575 &
                0.808 &
                0.219 &
               \cellcolor{blue!17} 0.270 &
                0.772 &
                0.763 &
                \cellcolor{blue!17}0.807 &
                523.05 &
                52.36 &
                 \\

            \midrule

            NRMS &
                G-KL &
                0.307 &
                0.482 &
                0.239 &
                0.608 &
                0.091 &
                0.544 &
                0.798 &
                \cellcolor{green!17}0.133 &
                \cellcolor{green!17}0.250 &
                0.790 &
                \cellcolor{green!17}0.779 &
                \cellcolor{green!17}0.789 &
                326.55 &
                30.15 &
                 \\

             &
                PM-2 &
                \cellcolor{blue!17}0.316 &
                0.483 &
                0.244 &
                0.598 &
                0.099 &
                0.544 &
                \cellcolor{orange!17}0.796 &
                \cellcolor{blue!17}0.150 &
                \cellcolor{green!17}0.250 &
                \cellcolor{orange!17}0.794 &
                \cellcolor{blue!17}0.776 &
                \cellcolor{green!17}0.789 &
                326.55 &
                23.86 &
                 \\

            &
                MMR &
                \cellcolor{orange!17}0.315 &
                0.477 &
                0.238 &
                0.616 &
                0.099 &
                \cellcolor{blue!17}0.571 &
                0.798 &
                \cellcolor{orange!17}0.204 &
                \cellcolor{blue!17}0.270 &
                0.790 &
                \cellcolor{orange!17}0.767 &
                \cellcolor{blue!17}0.807 &
                326.55 &
                52.36 &
                 \\

            \midrule

            \drdw{} &
                 &
                \cellcolor{green!17}0.374 &
                \cellcolor{green!17}0.407 &
                \cellcolor{blue!17}0.229 &
                \cellcolor{green!17}0.394 &
                0.107 &
                \cellcolor{orange!17}0.556 &
                0.798 &
                \cellcolor{green!17}0.133 &
                \cellcolor{green!17}0.250 &
                \cellcolor{blue!17}0.810 &
                \cellcolor{green!17}0.779 &
                \cellcolor{green!17}0.789 &
                \cellcolor{blue!17}2.81 &
                \cellcolor{orange!17}11.76 &
                \cellcolor{orange!17}0.554 \\

            \rdw{}~\cite{christoffel2015blockbusters} &
                 &
                0.222 &
                \cellcolor{blue!17}0.415 &
                \cellcolor{orange!17}0.235 &
                0.582 &
                0.080 &
                0.376 &
                0.840 &
                0.755 &
                0.856 &
                0.743 &
                0.439 &
                0.392 &
                \cellcolor{green!17}2.75 &
                \cellcolor{blue!17}1.27 &
                \cellcolor{green!17}0.565 \\

            RWE-D~\cite{paudel2021random} &
                 &
                0.256 &
                0.435 &
                \cellcolor{green!17}0.222 &
                \cellcolor{blue!17}0.443 &
                0.100 &
                0.372 &
                0.857 &
                0.802 & 
                \cellcolor{orange!17}0.842 &
                0.735 &
                0.377 & 
                \cellcolor{orange!17}0.433 &
                \cellcolor{orange!17}3.15 &
                \cellcolor{green!17}0.27 &
                \cellcolor{orange!17}0.554 \\

            \midrule

            Random &
                 & 
                0.180 &
                0.461 &
                0.256 &
                0.705 &
                \cellcolor{green!17}0.054 &
                0.366 &
                \cellcolor{green!17}0.756 &
                0.634 &
                0.873 &
                \cellcolor{green!17}0.842 &
                0.564 &
                0.346 &
                 &
                 &
                0.500 \\

            \bottomrule

            
        \end{tabular}

    }
    
    \Description{TBD}
    \label{tab:comparison}
    
\end{table*}

Lastly, we list the estimated energy cost to train models/create graphs (Train Cost) and to calculate the recommendations (Rec. Cost) in watt seconds.\footnote{We calculate the energy cost based on the overall power consumption (in watts for CPU and GPU components) and factor in the total runtime~\cite{vente2024clicks}. The values present upper-bound cost estimates.}
The evaluation of the entire experimental procedure is done using \textit{Informfully Recommenders}~\cite{heitz2025recommenders}, an extension of the Cornac Framework~\cite{salah2020cornac,truong2021exploring,truong2021multi}. 

\section{Results}
\label{sec:results}
Table~\ref{tab:comparison} shows the results of the RADio diversity, Gini, ILD, cost estimates, and AUC. 
We have added a row for Normative Target Values (NTV).
They show the scores that are used for ranking the top $3$ results in each column.
For \textit{norm-aware RADio metrics}, the values consider the underlying distributions of item properties in EB-NeRD and the user histories.
For \textit{traditional metrics}, the target values correspond to the best possible score under our chosen NTD.
Note that these are not the conventional bounds (e.g., 0 for Gini, 1 for ILD). 
Diversity metrics look at the top $20$ items of the recommendation list. 
AUC is determined based on the prediction score for all available items in the impression pool.

\textbf{Activation} compares article sentiment distributions, with higher values indicating a larger divergence between a user's recommendation list and the item pool.
\drdw{} together NRMS with PM-2 and MMR re-ranking results in the largest divergence.

\textbf{Calibration} compares the category and complexity distribution in the user history and their recommendation list.
A higher value signifies a greater deviation from the user's category/complexity preferences.
\drdw, \rdw, and LSTUR (with and without MMR) show the smallest divergence for category;
RWE-D, \drdw{}, \rdw{}, and LSTUR with MMR  have the smallest divergence for complexity.

\textbf{Fragmentation} quantifies the difference in story distribution across users.
A higher value indicates a greater variation in story chains, showing a more fragmented user base.
\drdw{}, RWE-D, and NPA with PM-2 provide the least fragmented user base.

\textbf{Alternative Voices} evaluates how the proportion of minority and majority perspectives in the recommended articles compares to their overall proportion in the available article pool.
A higher value indicates a greater disparity between the two perspectives.
The random baseline, NRMS, and NPA show the smallest disparity.

\textbf{Representation} measures the divergence in representing political parties.
A higher value indicates a larger difference between the recommendation list and the item pool.
LSTUR and NPA with MMR, NRMS with MMR, and \drdw{} show the largest divergence.

\textbf{Gini coefficients} quantify inequality in the target dimension. 
A higher value indicates larger inequality.
\drdw{} and the G-KL re-ranker outperform the base neural models across Gini of sentiment and party and achieve full NTV for all norm-relevant dimensions.
The random baseline achieved the best results for category Gini.

\textbf{ILD} calculates the average pairwise dissimilarity between items using cosine distance. 
A higher value means more dissimilarity (i.e., diversity) between items.
\drdw{} and baseline neural models combined with the G-KL re-ranker achieve the ILD value closest to the target ILD value, for the dimensions of sentiment and party.

\textbf{Computational Cost} lists the average watt seconds for calculating the recommendation list of $20$ items per user.
The training phase for the random walk models includes graph construction and graph augmentation.
The recommendation phase for graph-based algorithms consists of random walks, applying the distribution, and item re-ranking.
For re-rankers, we copied the training cost of the underlying model.\footnote{LSTUR, NPA, and NRMS ran on an Nvidia RTX 4090 GPU ($16,384$ cores at 2.5~GHz, 450~W power consumption). Random walk models used a CPU ($20$ cores at 3.9~GHz, 125~W power consumption).}
We see that the random walk models require one order of magnitude less energy for both the training as well as recommendation process when compared to the neural models.

\textbf{AUC} assesses model classification quality in predicting item clicks, with $0.5$ being the expected value of random guessing. 
We see \rdw has the highest AUC, followed by LSTUR in the second place, and a tie between \drdw{}, RWE-D, and NPA in third place.

\section{Discussion}
\label{sec:discussion}
There are four dimensions to our comparison: 1) normative RADio diversity, 2) traditional Gini/ILD diversity, 3) energy cost, and 4) AUC.
We look at the specific advantages of \drdw{} in comparison to 1) neural baseline models, 2) re-rankers with NTD-sampling, 3) other random walks, and 4) baselines.

Interpreting RADio metrics requires looking at the dataset first, as activation and representation metrics express a divergence between the underlying data and the recommendations.
EB-NeRD does not provide an equal split for party mentions nor sentiment.\footnote{Looking at the underlying distribution, a total of $2.66\%$ articles mention the governing parties and their supporters, $0.88\%$ the opposition parties, $0.41\%$ both governing and opposition parties, $16.96\%$ independent and foreign parties, and $79.09\%$ of articles have no political entities. The sentiment of $9.51\%$ articles is negative, $58.01\%$ somewhat negative, $30.01\%$ somewhat positive, and $2.48\%$ positive.}
We, therefore, aim for large divergence values when comparing to the underlying dataset.
Looking at the results for RADio in Table~\ref{tab:comparison}, we see 
\drdw{} achieves top scores for almost all metrics.
This shows that it can be used effectively to alter the divergence between recommendations and the underlying dataset.

For traditional diversity, the NTD in-process resampling of \drdw{} and the NTD post-processing for re-rankers allow targeting and boosting specific metrics across Gini as well as ILD for sentiment and party (matching the optimal NTV part of the NTD).
To us, this shows that a NTD-based sampling strategy as part of the recommendation pipeline is an effective way to control the diversity scores for the traditional metrics. 

While neural models with NTD re-ranking allow for combining AUC gains with excellent Gini and ILD performance, they require at the very least one order of magnitude more energy than the random walk algorithms.
Being cheaper to run, \drdw, \rdw, and RWE-D allow for more frequent updates of the recommendation lists.
This makes them ideal for the news domain, where there is a large number of items with a relatively short shelf life.

Lastly, we see that neural baselines and random walk models have more or less the same performance in terms of AUC.
While AUC values below $0.6$ indicate poor discriminatory power, it is important to emphasize that the news presents a difficult domain.
Reference AUC values for state-of-the-art algorithms are around this value for MIND~\cite{wu2020mind}, EB-NeRD~\cite{kruse2024eb}, and NeMig~\cite{iana2023nemig}. 

In summary, the results in Table~\ref{tab:comparison} show that \drdw{} has consistent scores in the top $3$ across all four dimensions.


\section{Limitations and Future Work}
\label{sec:limitations}

Since the dataset heavily influences the performance of recommender models and re-rankers, future work should better focus on examining the relationship between dataset characteristics and diversity outcomes. 
Second, further experiments could focus on better understanding the normative target distributions for multiple diversity dimensions on user engagement and satisfaction. 
Third, integrating personalized approaches to diversity is a promising direction for future work (i.e., what level of diversity is suitable for what type of user). 
To facilitate this, a user modeling module could be developed to enhance the system's ability to understand and adapt to user attributes, such as preferences for topics, political orientation, or levels of diversity.

Looking at future work,
user studies are required to understand what users perceive/accept as diverse and satisfying~\cite{heitz2024recommendations,vrijenhoek2024diversity}, what would be desirable approaches for visualizing diversity-driven recommended item lists (i.e., assessing the effectiveness of the graph coloring step of \drdw), and how normative-driven nudges could influence user acceptance and engagement with diversified content. 

\section{Conclusion}
\label{sec:conclusion}
In this paper, we present \drdw{}, a first operationalization of deliberative recommenders from the normative framework of~\citet{helberger2019democratic}. 
It is the best-performing solution when looking at the combined dimensions of RADio, Gini/ILD, energy cost, and AUC when compared to state-of-the-art neural models and random walks.
Furthermore, we provide a stand-alone version of the NTD sampler of D-RDW as a re-ranker for existing neural models.
The re-rankers allow for achieving more targeted results when added to the neural models across the traditional ILD and Gini diversity metrics.

The NTD sampling at the in- and post-processing steps is an effective and easily explainable approach to achieving a normative goal by virtue of adjusting the desired normative distribution based on enforcing distributions of article dimensions in news feeds.
This enables the input of editorial and journalistic values into recommendations.
We hope \drdw{} and random walks can become a cornerstone of exploring norm-aware NRSs that go beyond accuracy objectives by facilitating experiments with norms and values together with the assessment of their societal impact.

\balance

\begin{acks}
This work was partially funded by the Digital Society Initiative (DSI) of the University of Zurich (UZH) under a grant from the DSI Excellence Program, the Hasler Foundation, and the Swiss Federal Office of Communications (OFCOM). 
\end{acks}

\bibliographystyle{ACM-Reference-Format}
\bibliography{main}


\begin{thebibliography}{52}


\ifx \showCODEN    \undefined \def \showCODEN     #1{\unskip}     \fi
\ifx \showDOI      \undefined \def \showDOI       #1{#1}\fi
\ifx \showISBNx    \undefined \def \showISBNx     #1{\unskip}     \fi
\ifx \showISBNxiii \undefined \def \showISBNxiii  #1{\unskip}     \fi
\ifx \showISSN     \undefined \def \showISSN      #1{\unskip}     \fi
\ifx \showLCCN     \undefined \def \showLCCN      #1{\unskip}     \fi
\ifx \shownote     \undefined \def \shownote      #1{#1}          \fi
\ifx \showarticletitle \undefined \def \showarticletitle #1{#1}   \fi
\ifx \showURL      \undefined \def \showURL       {\relax}        \fi
\providecommand\bibfield[2]{#2}
\providecommand\bibinfo[2]{#2}
\providecommand\natexlab[1]{#1}
\providecommand\showeprint[2][]{arXiv:#2}

\bibitem[An et~al\mbox{.}(2019)]%
        {an2019lstur}
\bibfield{author}{\bibinfo{person}{Mingxiao An}, \bibinfo{person}{Fangzhao Wu}, \bibinfo{person}{Chuhan Wu}, \bibinfo{person}{Kun Zhang}, \bibinfo{person}{Zheng Liu}, {and} \bibinfo{person}{Xing Xie}.} \bibinfo{year}{2019}\natexlab{}.
\newblock \showarticletitle{Neural news recommendation with long-and short-term user representations}. In \bibinfo{booktitle}{\emph{Proceedings of the 57th annual meeting of the association for computational linguistics}}. \bibinfo{pages}{336--345}.
\newblock


\bibitem[Bachmann et~al\mbox{.}(2022)]%
        {bachmann2022defining}
\bibfield{author}{\bibinfo{person}{Philipp Bachmann}, \bibinfo{person}{Mark Eisenegger}, {and} \bibinfo{person}{Diana Ingenhoff}.} \bibinfo{year}{2022}\natexlab{}.
\newblock \showarticletitle{Defining and measuring news media quality: Comparing the content perspective and the audience perspective}.
\newblock \bibinfo{journal}{\emph{The International Journal of Press/Politics}} \bibinfo{volume}{27}, \bibinfo{number}{1} (\bibinfo{year}{2022}), \bibinfo{pages}{9--37}.
\newblock


\bibitem[Baden and Springer(2017)]%
        {baden2017conceptualizing}
\bibfield{author}{\bibinfo{person}{Christian Baden} {and} \bibinfo{person}{Nina Springer}.} \bibinfo{year}{2017}\natexlab{}.
\newblock \showarticletitle{Conceptualizing viewpoint diversity in news discourse}.
\newblock \bibinfo{journal}{\emph{Journalism}} \bibinfo{volume}{18}, \bibinfo{number}{2} (\bibinfo{year}{2017}), \bibinfo{pages}{176--194}.
\newblock


\bibitem[Baluja et~al\mbox{.}(2008)]%
        {baluja2008video}
\bibfield{author}{\bibinfo{person}{Shumeet Baluja}, \bibinfo{person}{Rohan Seth}, \bibinfo{person}{D. Sivakumar}, \bibinfo{person}{Yushi Jing}, \bibinfo{person}{Jay Yagnik}, \bibinfo{person}{Shankar Kumar}, \bibinfo{person}{Deepak Ravichandran}, {and} \bibinfo{person}{Mohamed Aly}.} \bibinfo{year}{2008}\natexlab{}.
\newblock \showarticletitle{Video suggestion and discovery for youtube: taking random walks through the view graph}. In \bibinfo{booktitle}{\emph{Proceedings of the 17th International Conference on World Wide Web}} (Beijing, China) \emph{(\bibinfo{series}{WWW '08})}. \bibinfo{publisher}{Association for Computing Machinery}, \bibinfo{address}{New York, NY, USA}, \bibinfo{pages}{895–904}.
\newblock
\showISBNx{9781605580852}
\urldef\tempurl%
\url{https://doi.org/10.1145/1367497.1367618}
\showDOI{\tempurl}


\bibitem[Bauer et~al\mbox{.}(2024)]%
        {bauer2024values}
\bibfield{author}{\bibinfo{person}{Christine Bauer}, \bibinfo{person}{Chandni Bagchi}, \bibinfo{person}{Olusanmi~A. Hundogan}, {and} \bibinfo{person}{Karin van Es}.} \bibinfo{year}{2024}\natexlab{}.
\newblock \showarticletitle{Where Are the Values? A Systematic Literature Review on News Recommender Systems}.
\newblock \bibinfo{journal}{\emph{ACM Trans. Recomm. Syst.}} \bibinfo{volume}{2}, \bibinfo{number}{3}, Article \bibinfo{articleno}{23} (\bibinfo{date}{jun} \bibinfo{year}{2024}), \bibinfo{numpages}{40}~pages.
\newblock
\urldef\tempurl%
\url{https://doi.org/10.1145/3654805}
\showDOI{\tempurl}


\bibitem[Bernstein et~al\mbox{.}(2021)]%
        {bernstein_et_al:DagMan.9.1.43}
\bibfield{author}{\bibinfo{person}{Abraham Bernstein}, \bibinfo{person}{Claes De~Vreese}, \bibinfo{person}{Natali Helberger}, \bibinfo{person}{Wolfgang Schulz}, \bibinfo{person}{Katharina Zweig}, \bibinfo{person}{Lucien Heitz}, \bibinfo{person}{Suzanne Tolmeijer}, {et~al\mbox{.}}} \bibinfo{year}{2021}\natexlab{}.
\newblock \showarticletitle{Diversity in News Recommendation (Dagstuhl Perspectives Workshop 19482)}.
\newblock \bibinfo{journal}{\emph{Dagstuhl Manifestos}} \bibinfo{volume}{9}, \bibinfo{number}{1} (\bibinfo{year}{2021}), \bibinfo{pages}{43--61}.
\newblock
\showISSN{2193-2433}
\urldef\tempurl%
\url{https://doi.org/10.4230/DagMan.9.1.43}
\showDOI{\tempurl}


\bibitem[Carbonell and Goldstein(1998)]%
        {carbonell1998use}
\bibfield{author}{\bibinfo{person}{Jaime Carbonell} {and} \bibinfo{person}{Jade Goldstein}.} \bibinfo{year}{1998}\natexlab{}.
\newblock \showarticletitle{The use of MMR, diversity-based reranking for reordering documents and producing summaries}. In \bibinfo{booktitle}{\emph{Proceedings of the 21st annual international ACM SIGIR conference on Research and development in information retrieval}}. \bibinfo{pages}{335--336}.
\newblock


\bibitem[Chakraborty et~al\mbox{.}(2017)]%
        {chakraborty2017optimizing}
\bibfield{author}{\bibinfo{person}{Abhijnan Chakraborty}, \bibinfo{person}{Saptarshi Ghosh}, \bibinfo{person}{Niloy Ganguly}, {and} \bibinfo{person}{Krishna~P Gummadi}.} \bibinfo{year}{2017}\natexlab{}.
\newblock \showarticletitle{Optimizing the recency-relevancy trade-off in online news recommendations}. In \bibinfo{booktitle}{\emph{Proceedings of the 26th International Conference on World Wide Web}}. \bibinfo{pages}{837--846}.
\newblock


\bibitem[Christoffel et~al\mbox{.}(2015)]%
        {christoffel2015blockbusters}
\bibfield{author}{\bibinfo{person}{Fabian Christoffel}, \bibinfo{person}{Bibek Paudel}, \bibinfo{person}{Chris Newell}, {and} \bibinfo{person}{Abraham Bernstein}.} \bibinfo{year}{2015}\natexlab{}.
\newblock \showarticletitle{Blockbusters and Wallflowers: Accurate, Diverse, and Scalable Recommendations with Random Walks}. In \bibinfo{booktitle}{\emph{Proceedings of the 9th ACM Conference on Recommender Systems}} (Vienna, Austria) \emph{(\bibinfo{series}{RecSys '15})}. \bibinfo{publisher}{Association for Computing Machinery}, \bibinfo{address}{New York, NY, USA}, \bibinfo{pages}{163–170}.
\newblock
\showISBNx{9781450336925}
\urldef\tempurl%
\url{https://doi.org/10.1145/2792838.2800180}
\showDOI{\tempurl}


\bibitem[Dang and Croft(2012)]%
        {dang2012diversity}
\bibfield{author}{\bibinfo{person}{Van Dang} {and} \bibinfo{person}{W~Bruce Croft}.} \bibinfo{year}{2012}\natexlab{}.
\newblock \showarticletitle{Diversity by proportionality: an election-based approach to search result diversification}. In \bibinfo{booktitle}{\emph{Proceedings of the 35th international ACM SIGIR conference on Research and development in information retrieval}}. \bibinfo{pages}{65--74}.
\newblock


\bibitem[De~Gemmis et~al\mbox{.}(2015)]%
        {de2015investigation}
\bibfield{author}{\bibinfo{person}{Marco De~Gemmis}, \bibinfo{person}{Pasquale Lops}, \bibinfo{person}{Giovanni Semeraro}, {and} \bibinfo{person}{Cataldo Musto}.} \bibinfo{year}{2015}\natexlab{}.
\newblock \showarticletitle{An investigation on the serendipity problem in recommender systems}.
\newblock \bibinfo{journal}{\emph{Information Processing \& Management}} \bibinfo{volume}{51}, \bibinfo{number}{5} (\bibinfo{year}{2015}), \bibinfo{pages}{695--717}.
\newblock


\bibitem[Eksombatchai et~al\mbox{.}(2018)]%
        {eksombatchai2018pixie}
\bibfield{author}{\bibinfo{person}{Chantat Eksombatchai}, \bibinfo{person}{Pranav Jindal}, \bibinfo{person}{Jerry~Zitao Liu}, \bibinfo{person}{Yuchen Liu}, \bibinfo{person}{Rahul Sharma}, \bibinfo{person}{Charles Sugnet}, \bibinfo{person}{Mark Ulrich}, {and} \bibinfo{person}{Jure Leskovec}.} \bibinfo{year}{2018}\natexlab{}.
\newblock \showarticletitle{Pixie: A system for recommending 3+ billion items to 200+ million users in real-time}. In \bibinfo{booktitle}{\emph{Proceedings of the 2018 world wide web conference}}. \bibinfo{pages}{1775--1784}.
\newblock


\bibitem[Ferrari~Dacrema et~al\mbox{.}(2019)]%
        {ferrari2019we}
\bibfield{author}{\bibinfo{person}{Maurizio Ferrari~Dacrema}, \bibinfo{person}{Paolo Cremonesi}, {and} \bibinfo{person}{Dietmar Jannach}.} \bibinfo{year}{2019}\natexlab{}.
\newblock \showarticletitle{Are we really making much progress? A worrying analysis of recent neural recommendation approaches}. In \bibinfo{booktitle}{\emph{Proceedings of the 13th ACM conference on recommender systems}}. \bibinfo{pages}{101--109}.
\newblock


\bibitem[Heitz(2023)]%
        {heitz2023classification}
\bibfield{author}{\bibinfo{person}{Lucien Heitz}.} \bibinfo{year}{2023}\natexlab{}.
\newblock \showarticletitle{Classification of Normative Recommender Systems}. In \bibinfo{booktitle}{\emph{Proceedings of the First Workshop on the Normative Design and Evaluation of Recommender Systems}}.
\newblock


\bibitem[Heitz et~al\mbox{.}(2024a)]%
        {heitz2024informfully}
\bibfield{author}{\bibinfo{person}{Lucien Heitz}, \bibinfo{person}{Julian~A Croci}, \bibinfo{person}{Madhav Sachdeva}, {and} \bibinfo{person}{Abraham Bernstein}.} \bibinfo{year}{2024}\natexlab{a}.
\newblock \showarticletitle{Informfully - Research Platform for Reproducible User Studies}. In \bibinfo{booktitle}{\emph{Proceedings of the 18th ACM Conference on Recommender Systems}}.
\newblock


\bibitem[Heitz et~al\mbox{.}(2024b)]%
        {heitz2024recommendations}
\bibfield{author}{\bibinfo{person}{Lucien Heitz}, \bibinfo{person}{Oana Inel}, {and} \bibinfo{person}{Sanne Vrijenhoek}.} \bibinfo{year}{2024}\natexlab{b}.
\newblock \showarticletitle{Recommendations for the Recommenders: Reflections on Prioritizing Diversity in the RecSys Challenge}. In \bibinfo{booktitle}{\emph{Proceedings of the Recommender Systems Challenge 2024}}. \bibinfo{pages}{22--26}.
\newblock


\bibitem[Heitz et~al\mbox{.}(2025)]%
        {heitz2025recommenders}
\bibfield{author}{\bibinfo{person}{Lucien Heitz}, \bibinfo{person}{Runze Li}, \bibinfo{person}{Oana Inel}, {and} \bibinfo{person}{Abraham Bernstein}.} \bibinfo{year}{2025}\natexlab{}.
\newblock \showarticletitle{Informfully Recommenders – Reproducibility Framework for Diversity-aware Intra-session Recommendations}. In \bibinfo{booktitle}{\emph{Proceedings of the 19th ACM Conference on Recommender Systems}}.
\newblock


\bibitem[Heitz et~al\mbox{.}(2023)]%
        {heitz2023deliberative}
\bibfield{author}{\bibinfo{person}{Lucien Heitz}, \bibinfo{person}{Juliane~A Lischka}, \bibinfo{person}{Rana Abdullah}, \bibinfo{person}{Laura Laugwitz}, \bibinfo{person}{Hendrik Meyer}, {and} \bibinfo{person}{Abraham Bernstein}.} \bibinfo{year}{2023}\natexlab{}.
\newblock \showarticletitle{Deliberative Diversity for News Recommendations: Operationalization and Experimental User Study}. In \bibinfo{booktitle}{\emph{Proceedings of the 17th ACM Conference on Recommender Systems}}. \bibinfo{pages}{813--819}.
\newblock


\bibitem[Heitz et~al\mbox{.}(2022)]%
        {heitz2022benefits}
\bibfield{author}{\bibinfo{person}{Lucien Heitz}, \bibinfo{person}{Juliane~A Lischka}, \bibinfo{person}{Alena Birrer}, \bibinfo{person}{Bibek Paudel}, \bibinfo{person}{Suzanne Tolmeijer}, \bibinfo{person}{Laura Laugwitz}, {and} \bibinfo{person}{Abraham Bernstein}.} \bibinfo{year}{2022}\natexlab{}.
\newblock \showarticletitle{Benefits of diverse news recommendations for democracy: A user study}.
\newblock \bibinfo{journal}{\emph{Digital Journalism}} \bibinfo{volume}{10}, \bibinfo{number}{10} (\bibinfo{year}{2022}), \bibinfo{pages}{1710--1730}.
\newblock


\bibitem[Heitz et~al\mbox{.}(2024c)]%
        {heitz2024idea}
\bibfield{author}{\bibinfo{person}{Lucien Heitz}, \bibinfo{person}{Nicolas Mattis}, \bibinfo{person}{Oana Inel}, {and} \bibinfo{person}{Wouter van Atteveldt}.} \bibinfo{year}{2024}\natexlab{c}.
\newblock \showarticletitle{IDEA – Informfully Dataset with Enhanced Attributes}. In \bibinfo{booktitle}{\emph{Proceedings of the Second Workshop on the Normative Design and Evaluation of Recommender Systems}}.
\newblock


\bibitem[Helberger(2019)]%
        {helberger2019democratic}
\bibfield{author}{\bibinfo{person}{Natali Helberger}.} \bibinfo{year}{2019}\natexlab{}.
\newblock \showarticletitle{On the Democratic Role of News Recommenders}.
\newblock \bibinfo{journal}{\emph{Digital Journalism}} \bibinfo{volume}{7}, \bibinfo{number}{8} (\bibinfo{year}{2019}), \bibinfo{pages}{993--1012}.
\newblock


\bibitem[Helberger et~al\mbox{.}(2018)]%
        {helberger2018exposure}
\bibfield{author}{\bibinfo{person}{Natali Helberger}, \bibinfo{person}{Kari Karppinen}, {and} \bibinfo{person}{Lucia D’acunto}.} \bibinfo{year}{2018}\natexlab{}.
\newblock \showarticletitle{Exposure diversity as a design principle for recommender systems}.
\newblock \bibinfo{journal}{\emph{Information, Communication \& Society}} \bibinfo{volume}{21}, \bibinfo{number}{2} (\bibinfo{year}{2018}), \bibinfo{pages}{191--207}.
\newblock


\bibitem[Iana et~al\mbox{.}(2023)]%
        {iana2023nemig}
\bibfield{author}{\bibinfo{person}{Andreea Iana}, \bibinfo{person}{Mehwish Alam}, \bibinfo{person}{Alexander Grote}, \bibinfo{person}{Nevena Nikolajevic}, \bibinfo{person}{Katharina Ludwig}, \bibinfo{person}{Philipp M{\"u}ller}, \bibinfo{person}{Christof Weinhardt}, {and} \bibinfo{person}{Heiko Paulheim}.} \bibinfo{year}{2023}\natexlab{}.
\newblock \showarticletitle{NeMig-A Bilingual News Collection and Knowledge Graph about Migration}.
\newblock  (\bibinfo{year}{2023}).
\newblock


\bibitem[Kierstead(2000)]%
        {kierstead2000simple}
\bibfield{author}{\bibinfo{person}{Hal~A Kierstead}.} \bibinfo{year}{2000}\natexlab{}.
\newblock \showarticletitle{A simple competitive graph coloring algorithm}.
\newblock \bibinfo{journal}{\emph{Journal of Combinatorial Theory, Series B}} \bibinfo{volume}{78}, \bibinfo{number}{1} (\bibinfo{year}{2000}), \bibinfo{pages}{57--68}.
\newblock


\bibitem[Kruse et~al\mbox{.}(2024a)]%
        {kruse2024eb}
\bibfield{author}{\bibinfo{person}{Johannes Kruse}, \bibinfo{person}{Kasper Lindskow}, \bibinfo{person}{Saikishore Kalloori}, \bibinfo{person}{Marco Polignano}, \bibinfo{person}{Claudio Pomo}, \bibinfo{person}{Abhishek Srivastava}, \bibinfo{person}{Anshuk Uppal}, \bibinfo{person}{Michael~Riis Andersen}, {and} \bibinfo{person}{Jes Frellsen}.} \bibinfo{year}{2024}\natexlab{a}.
\newblock \showarticletitle{EB-NeRD a large-scale dataset for news recommendation}. In \bibinfo{booktitle}{\emph{Proceedings of the Recommender Systems Challenge 2024}}. \bibinfo{pages}{1--11}.
\newblock


\bibitem[Kruse et~al\mbox{.}(2024b)]%
        {kruse2024recsys}
\bibfield{author}{\bibinfo{person}{Johannes Kruse}, \bibinfo{person}{Kasper Lindskow}, \bibinfo{person}{Saikishore Kalloori}, \bibinfo{person}{Marco Polignano}, \bibinfo{person}{Claudio Pomo}, \bibinfo{person}{Abhishek Srivastava}, \bibinfo{person}{Anshuk Uppal}, \bibinfo{person}{Michael~Riis Andersen}, {and} \bibinfo{person}{Jes Frellsen}.} \bibinfo{year}{2024}\natexlab{b}.
\newblock \showarticletitle{RecSys Challenge 2024: Balancing Accuracy and Editorial Values in News Recommendations}. In \bibinfo{booktitle}{\emph{Proceedings of the 18th ACM Conference on Recommender Systems}}. \bibinfo{pages}{1195--1199}.
\newblock


\bibitem[Liu et~al\mbox{.}(2012)]%
        {liu2012solving}
\bibfield{author}{\bibinfo{person}{Jian-Guo Liu}, \bibinfo{person}{Kerui Shi}, {and} \bibinfo{person}{Qiang Guo}.} \bibinfo{year}{2012}\natexlab{}.
\newblock \showarticletitle{Solving the accuracy-diversity dilemma via directed random walks}.
\newblock \bibinfo{journal}{\emph{Physical Review E}} \bibinfo{volume}{85}, \bibinfo{number}{1} (\bibinfo{year}{2012}), \bibinfo{pages}{016118}.
\newblock


\bibitem[Liu and Yang(2008)]%
        {liu2008eigenrank}
\bibfield{author}{\bibinfo{person}{Nathan~N Liu} {and} \bibinfo{person}{Qiang Yang}.} \bibinfo{year}{2008}\natexlab{}.
\newblock \showarticletitle{Eigenrank: a ranking-oriented approach to collaborative filtering}. In \bibinfo{booktitle}{\emph{Proceedings of the 31st annual international ACM SIGIR conference on Research and development in information retrieval}}. \bibinfo{pages}{83--90}.
\newblock


\bibitem[Modre et~al\mbox{.}(2023)]%
        {modre2023value}
\bibfield{author}{\bibinfo{person}{Laura Modre}, \bibinfo{person}{Julia Neidhardt}, {and} \bibinfo{person}{Irina Nalis}.} \bibinfo{year}{2023}\natexlab{}.
\newblock \showarticletitle{Value-Based Nudging in News Recommender Systems-Results From an Experimental User Study.}. In \bibinfo{booktitle}{\emph{NORMalize@ RecSys}}.
\newblock


\bibitem[Paudel and Bernstein(2021)]%
        {paudel2021random}
\bibfield{author}{\bibinfo{person}{Bibek Paudel} {and} \bibinfo{person}{Abraham Bernstein}.} \bibinfo{year}{2021}\natexlab{}.
\newblock \showarticletitle{Random walks with erasure: Diversifying personalized recommendations on social and information networks}. In \bibinfo{booktitle}{\emph{Proceedings of the Web Conference 2021}}. \bibinfo{pages}{2046--2057}.
\newblock


\bibitem[Paudel et~al\mbox{.}(2016)]%
        {paudel2016updatable}
\bibfield{author}{\bibinfo{person}{Bibek Paudel}, \bibinfo{person}{Fabian Christoffel}, \bibinfo{person}{Chris Newell}, {and} \bibinfo{person}{Abraham Bernstein}.} \bibinfo{year}{2016}\natexlab{}.
\newblock \showarticletitle{Updatable, accurate, diverse, and scalable recommendations for interactive applications}.
\newblock \bibinfo{journal}{\emph{ACM Transactions on Interactive Intelligent Systems (TiiS)}} \bibinfo{volume}{7}, \bibinfo{number}{1} (\bibinfo{year}{2016}), \bibinfo{pages}{1--34}.
\newblock


\bibitem[Petersen et~al\mbox{.}(2021)]%
        {petersen2021post}
\bibfield{author}{\bibinfo{person}{Felix Petersen}, \bibinfo{person}{Debarghya Mukherjee}, \bibinfo{person}{Yuekai Sun}, {and} \bibinfo{person}{Mikhail Yurochkin}.} \bibinfo{year}{2021}\natexlab{}.
\newblock \showarticletitle{Post-processing for individual fairness}.
\newblock \bibinfo{journal}{\emph{Advances in Neural Information Processing Systems}}  \bibinfo{volume}{34} (\bibinfo{year}{2021}), \bibinfo{pages}{25944--25955}.
\newblock


\bibitem[Provost and Fawcett(2001)]%
        {provost2001robust}
\bibfield{author}{\bibinfo{person}{Foster Provost} {and} \bibinfo{person}{Tom Fawcett}.} \bibinfo{year}{2001}\natexlab{}.
\newblock \showarticletitle{Robust classification for imprecise environments}.
\newblock \bibinfo{journal}{\emph{Machine learning}}  \bibinfo{volume}{42} (\bibinfo{year}{2001}), \bibinfo{pages}{203--231}.
\newblock


\bibitem[Rao et~al\mbox{.}(2013)]%
        {rao2013taxonomy}
\bibfield{author}{\bibinfo{person}{Junyang Rao}, \bibinfo{person}{Aixia Jia}, \bibinfo{person}{Yansong Feng}, {and} \bibinfo{person}{Dongyan Zhao}.} \bibinfo{year}{2013}\natexlab{}.
\newblock \showarticletitle{Taxonomy based personalized news recommendation: Novelty and diversity}. In \bibinfo{booktitle}{\emph{International Conference on Web Information Systems Engineering}}. Springer, \bibinfo{pages}{209--218}.
\newblock


\bibitem[Salah et~al\mbox{.}(2020)]%
        {salah2020cornac}
\bibfield{author}{\bibinfo{person}{Aghiles Salah}, \bibinfo{person}{Quoc-Tuan Truong}, {and} \bibinfo{person}{Hady~W Lauw}.} \bibinfo{year}{2020}\natexlab{}.
\newblock \showarticletitle{Cornac: A Comparative Framework for Multimodal Recommender Systems}.
\newblock \bibinfo{journal}{\emph{Journal of Machine Learning Research}} \bibinfo{volume}{21}, \bibinfo{number}{95} (\bibinfo{year}{2020}), \bibinfo{pages}{1--5}.
\newblock


\bibitem[Sargeant et~al\mbox{.}(2022)]%
        {sargeant2022spotlight}
\bibfield{author}{\bibinfo{person}{Holli Sargeant}, \bibinfo{person}{Eliska Pirkova}, \bibinfo{person}{Matthias~C Kettemann}, \bibinfo{person}{Marlena Wisniak}, \bibinfo{person}{Martin Scheinin}, \bibinfo{person}{Emmi Bevensee}, \bibinfo{person}{Katie Pentney}, \bibinfo{person}{Lorna Woods}, \bibinfo{person}{Lucien Heitz}, \bibinfo{person}{Bojana Kostic}, {et~al\mbox{.}}} \bibinfo{year}{2022}\natexlab{}.
\newblock \bibinfo{booktitle}{\emph{Spotlight on Artificial Intelligence and Freedom of Expression: A Policy Manual}}.
\newblock \bibinfo{publisher}{Organization for Security and Co-operation in Europe}.
\newblock


\bibitem[Steck(2018)]%
        {steck2018calibrated}
\bibfield{author}{\bibinfo{person}{Harald Steck}.} \bibinfo{year}{2018}\natexlab{}.
\newblock \showarticletitle{Calibrated recommendations}. In \bibinfo{booktitle}{\emph{Proceedings of the 12th ACM conference on recommender systems}}. \bibinfo{pages}{154--162}.
\newblock


\bibitem[Tintarev et~al\mbox{.}(2024)]%
        {tintarev2024measuring}
\bibfield{author}{\bibinfo{person}{Nava Tintarev}, \bibinfo{person}{Bart~P Knijnenburg}, {and} \bibinfo{person}{Martijn~C Willemsen}.} \bibinfo{year}{2024}\natexlab{}.
\newblock \showarticletitle{Measuring the benefit of increased transparency and control in news recommendation}.
\newblock \bibinfo{journal}{\emph{Ai Magazine}} \bibinfo{volume}{45}, \bibinfo{number}{2} (\bibinfo{year}{2024}), \bibinfo{pages}{212--226}.
\newblock


\bibitem[Truong et~al\mbox{.}(2021a)]%
        {truong2021multi}
\bibfield{author}{\bibinfo{person}{Quoc-Tuan Truong}, \bibinfo{person}{Aghiles Salah}, {and} \bibinfo{person}{Hady Lauw}.} \bibinfo{year}{2021}\natexlab{a}.
\newblock \showarticletitle{Multi-modal recommender systems: Hands-on exploration}. In \bibinfo{booktitle}{\emph{Fifteenth ACM Conference on Recommender Systems}}. \bibinfo{pages}{834--837}.
\newblock


\bibitem[Truong et~al\mbox{.}(2021b)]%
        {truong2021exploring}
\bibfield{author}{\bibinfo{person}{Quoc-Tuan Truong}, \bibinfo{person}{Aghiles Salah}, \bibinfo{person}{Thanh-Binh Tran}, \bibinfo{person}{Jingyao Guo}, {and} \bibinfo{person}{Hady~W Lauw}.} \bibinfo{year}{2021}\natexlab{b}.
\newblock \showarticletitle{Exploring Cross-Modality Utilization in Recommender Systems}.
\newblock \bibinfo{journal}{\emph{IEEE Internet Computing}} (\bibinfo{year}{2021}).
\newblock


\bibitem[Vargas et~al\mbox{.}(2014)]%
        {vargas2014coverage}
\bibfield{author}{\bibinfo{person}{Sa{\'u}l Vargas}, \bibinfo{person}{Linas Baltrunas}, \bibinfo{person}{Alexandros Karatzoglou}, {and} \bibinfo{person}{Pablo Castells}.} \bibinfo{year}{2014}\natexlab{}.
\newblock \showarticletitle{Coverage, redundancy and size-awareness in genre diversity for recommender systems}. In \bibinfo{booktitle}{\emph{Proceedings of the 8th ACM Conference on Recommender systems}}. \bibinfo{pages}{209--216}.
\newblock


\bibitem[Vente et~al\mbox{.}(2024)]%
        {vente2024clicks}
\bibfield{author}{\bibinfo{person}{Tobias Vente}, \bibinfo{person}{Lukas Wegmeth}, \bibinfo{person}{Alan Said}, {and} \bibinfo{person}{Joeran Beel}.} \bibinfo{year}{2024}\natexlab{}.
\newblock \showarticletitle{From clicks to carbon: The environmental toll of recommender systems}. In \bibinfo{booktitle}{\emph{Proceedings of the 18th ACM Conference on Recommender Systems}}. \bibinfo{pages}{580--590}.
\newblock


\bibitem[Vrijenhoek et~al\mbox{.}(2022)]%
        {vrijenhoek2022radio}
\bibfield{author}{\bibinfo{person}{Sanne Vrijenhoek}, \bibinfo{person}{Gabriel B{\'e}n{\'e}dict}, \bibinfo{person}{Mateo Gutierrez~Granada}, \bibinfo{person}{Daan Odijk}, {and} \bibinfo{person}{Maarten De~Rijke}.} \bibinfo{year}{2022}\natexlab{}.
\newblock \showarticletitle{Radio--rank-aware divergence metrics to measure normative diversity in news recommendations}. In \bibinfo{booktitle}{\emph{Proceedings of the 16th ACM Conference on Recommender Systems}}. \bibinfo{pages}{208--219}.
\newblock


\bibitem[Vrijenhoek et~al\mbox{.}(2024)]%
        {vrijenhoek2024diversity}
\bibfield{author}{\bibinfo{person}{Sanne Vrijenhoek}, \bibinfo{person}{Savvina Daniil}, \bibinfo{person}{Jorden Sandel}, {and} \bibinfo{person}{Laura Hollink}.} \bibinfo{year}{2024}\natexlab{}.
\newblock \showarticletitle{Diversity of What? On the Different Conceptualizations of Diversity in Recommender Systems}. In \bibinfo{booktitle}{\emph{Proceedings of the 2024 ACM Conference on Fairness, Accountability, and Transparency}} (Rio de Janeiro, Brazil) \emph{(\bibinfo{series}{FAccT '24})}. \bibinfo{publisher}{Association for Computing Machinery}, \bibinfo{address}{New York, NY, USA}, \bibinfo{pages}{573–584}.
\newblock
\showISBNx{9798400704505}
\urldef\tempurl%
\url{https://doi.org/10.1145/3630106.3658926}
\showDOI{\tempurl}


\bibitem[Vrijenhoek et~al\mbox{.}(2023)]%
        {vrijenhoek2023normalize}
\bibfield{author}{\bibinfo{person}{Sanne Vrijenhoek}, \bibinfo{person}{Lien Michiels}, \bibinfo{person}{Johannes Kruse}, \bibinfo{person}{Alain Starke}, \bibinfo{person}{Nava Tintarev}, {and} \bibinfo{person}{Jordi Viader~Guerrero}.} \bibinfo{year}{2023}\natexlab{}.
\newblock \showarticletitle{Normalize: The first workshop on normative design and evaluation of recommender systems}. In \bibinfo{booktitle}{\emph{Proceedings of the 17th ACM Conference on Recommender Systems}}. \bibinfo{pages}{1252--1254}.
\newblock


\bibitem[Wan et~al\mbox{.}(2023)]%
        {wan2023processing}
\bibfield{author}{\bibinfo{person}{Mingyang Wan}, \bibinfo{person}{Daochen Zha}, \bibinfo{person}{Ninghao Liu}, {and} \bibinfo{person}{Na Zou}.} \bibinfo{year}{2023}\natexlab{}.
\newblock \showarticletitle{In-processing modeling techniques for machine learning fairness: A survey}.
\newblock \bibinfo{journal}{\emph{ACM Transactions on Knowledge Discovery from Data}} \bibinfo{volume}{17}, \bibinfo{number}{3} (\bibinfo{year}{2023}), \bibinfo{pages}{1--27}.
\newblock


\bibitem[Wang et~al\mbox{.}(2018)]%
        {wang2018rndm}
\bibfield{author}{\bibinfo{person}{Mengsha Wang}, \bibinfo{person}{Yingyuan Xiao}, \bibinfo{person}{Wenguang Zheng}, {and} \bibinfo{person}{Xu Jiao}.} \bibinfo{year}{2018}\natexlab{}.
\newblock \showarticletitle{RNDM: A random walk method for music recommendation by considering novelty, diversity, and mainstream}. In \bibinfo{booktitle}{\emph{2018 IEEE 30th International Conference on Tools with Artificial Intelligence (ICTAI)}}. IEEE, \bibinfo{pages}{177--183}.
\newblock


\bibitem[Wu et~al\mbox{.}(2019a)]%
        {wu2019npa}
\bibfield{author}{\bibinfo{person}{Chuhan Wu}, \bibinfo{person}{Fangzhao Wu}, \bibinfo{person}{Mingxiao An}, \bibinfo{person}{Jianqiang Huang}, \bibinfo{person}{Yongfeng Huang}, {and} \bibinfo{person}{Xing Xie}.} \bibinfo{year}{2019}\natexlab{a}.
\newblock \showarticletitle{NPA: neural news recommendation with personalized attention}. In \bibinfo{booktitle}{\emph{Proceedings of the 25th ACM SIGKDD international conference on knowledge discovery \& data mining}}. \bibinfo{pages}{2576--2584}.
\newblock


\bibitem[Wu et~al\mbox{.}(2019b)]%
        {wu2019nrms}
\bibfield{author}{\bibinfo{person}{Chuhan Wu}, \bibinfo{person}{Fangzhao Wu}, \bibinfo{person}{Suyu Ge}, \bibinfo{person}{Tao Qi}, \bibinfo{person}{Yongfeng Huang}, {and} \bibinfo{person}{Xing Xie}.} \bibinfo{year}{2019}\natexlab{b}.
\newblock \showarticletitle{Neural news recommendation with multi-head self-attention}. In \bibinfo{booktitle}{\emph{Proceedings of the 2019 conference on empirical methods in natural language processing and the 9th international joint conference on natural language processing (EMNLP-IJCNLP)}}. \bibinfo{pages}{6389--6394}.
\newblock


\bibitem[Wu et~al\mbox{.}(2020b)]%
        {wu2020sentirec}
\bibfield{author}{\bibinfo{person}{Chuhan Wu}, \bibinfo{person}{Fangzhao Wu}, \bibinfo{person}{Tao Qi}, {and} \bibinfo{person}{Yongfeng Huang}.} \bibinfo{year}{2020}\natexlab{b}.
\newblock \showarticletitle{Sentirec: Sentiment diversity-aware neural news recommendation}. In \bibinfo{booktitle}{\emph{Proceedings of the 1st conference of the Asia-Pacific chapter of the association for computational linguistics and the 10th international joint conference on natural language processing}}. \bibinfo{pages}{44--53}.
\newblock


\bibitem[Wu et~al\mbox{.}(2020a)]%
        {wu2020mind}
\bibfield{author}{\bibinfo{person}{Fangzhao Wu}, \bibinfo{person}{Ying Qiao}, \bibinfo{person}{Jiun-Hung Chen}, \bibinfo{person}{Chuhan Wu}, \bibinfo{person}{Tao Qi}, \bibinfo{person}{Jianxun Lian}, \bibinfo{person}{Danyang Liu}, \bibinfo{person}{Xing Xie}, \bibinfo{person}{Jianfeng Gao}, \bibinfo{person}{Winnie Wu}, {et~al\mbox{.}}} \bibinfo{year}{2020}\natexlab{a}.
\newblock \showarticletitle{Mind: A large-scale dataset for news recommendation}. In \bibinfo{booktitle}{\emph{Proceedings of the 58th annual meeting of the association for computational linguistics}}. \bibinfo{pages}{3597--3606}.
\newblock


\bibitem[Yu et~al\mbox{.}(2019)]%
        {yu2019recommendation}
\bibfield{author}{\bibinfo{person}{Ting Yu}, \bibinfo{person}{Junpeng Guo}, \bibinfo{person}{Wenhua Li}, \bibinfo{person}{Harry~Jiannan Wang}, {and} \bibinfo{person}{Ling Fan}.} \bibinfo{year}{2019}\natexlab{}.
\newblock \showarticletitle{Recommendation with diversity: An adaptive trust-aware model}.
\newblock \bibinfo{journal}{\emph{Decision Support Systems}}  \bibinfo{volume}{123} (\bibinfo{year}{2019}), \bibinfo{pages}{113073}.
\newblock


\end{thebibliography}

\end{document}